# THE EFFECTS OF SMOOTHING ON THE STATISTICAL PROPERTIES OF LARGE–SCALE COSMIC FIELDS

by

Francis BERNARDEAU

CITA, 60 St George St., Toronto, Ontario, Canada M5S 1A1

## ABSTRACT

It has been shown that the large–scale correlation functions of the density field (and velocity divergence field) follow a specific hierarchy in the quasilinear regime and for Gaussian initial conditions (Bernardeau 1992). The exact relationships between the cumulants of the probability distribution functions (the so-called $S_p$ parameters) are however sensitive to the smoothing window function applied to the fields. In this paper, I present a method to derive the whole series of the $S_p$ parameters when the density field is smoothed with a top–hat window function. The results are valid for any power spectrum and any cosmological parameters. Similar calculations are presented for the velocity divergence field.

The resulting shapes of the one–point probability distribution functions of the cosmic density and the velocity divergence fields are given as a function of the power spectrum and $\Omega$. Simple analytical fits are proposed when the index of the power spectrum is $-1$. Comparisons with numerical simulations prove these analytical results to be extremely accurate.

## 1. Introduction

In most cosmological models, the large–scale structures of the Universe are thought to form by gravitational instabilities from small density fluctuations. In the mildly nonlinear regime, where the density fluctuations are still small, the mass distribution and motion of the Universe can reliably be described by the behavior of a pressureless fluid in an expanding Universe. It is then reasonable to expect perturbation theories to give a good description of the statistical properties of the large–scale cosmic fields. The linear theory has been widely evoked for the evolution of the rms density fluctuation and it has been also used for the velocity density relationship at large scales. However other large–scale features escape the first order perturbation theory, such as, for instance, the skewness of the density distribution function for Gaussian initial conditions. Indeed it is given by the second order perturbation theory (Peebles 1980). Actually, the characterization of the large–scale matter distribution and motion involve many more quantities than the simple rms density or velocity fluctuation. In the mildly nonlinear regime the shape the one-point density probability distribution function significantly departs from a Gaussian distribution. This evolution can be seen through the emergence of non-zero cumulants (of order greater than two) for this distribution, the skewness being the first one.

The assumption of Gaussian initial conditions is crucial for the derivation of such quantities and in the following, I will assume that the initial fluctuations follow Gaussian statistics. The original derivation of the skewness by Peebles has been improved since to take into account the effects of filtering (that in practice cannot be avoided) with numerical (Goroff et al. 1986, Grinstein & Wise 1987) or analytical (Juszkiewicz, Bouchet & Colombi 1993a, Bernardeau 1994b) calculations and the effects of redshift space distorsion (Bouchet et al. 1992) as well as its $\Omega$ and $\Lambda$ dependences. On the other hand, Fry (1984), Fry, Melott & Shandarin (1993) explored the geometrical dependences of the three-point correlation function given by these perturbative calculations. Similar calculations for the divergence of the velocity field have also been done (Bernardeau et al. 1994, Bernardeau 1994b) leading to new tools to constrain the cosmological parameters from the statistics of the large–scale cosmic velocity field. The kurtosis, obtained from the fourth moment of the distribution function, has been as well examined in details with perturbation theory. This quantity was originally derived by Fry (1984) and by Goroff et al. (1986) from third order perturbation theory. The effects of smoothing with a top hat window function on this quantity have been calculated analytically by Bernardeau (1994b) for both the density and the velocity divergence. For all these results the perturbation theory proved its efficiency when it has been compared with numerical simulations (previous reference and Juszkiewicz et al. 1993b).

However, up to now, only the skewness and the kurtosis were known when the smoothing effects are taken into account, whereas Bernardeau (1992) was able to derive the full hierarchy of the large–scale cumulants when the smoothing effects are neglected. The knowledge of such a hierarchy makes possible the derivation of the shape of the one–point probability distribution function of the density or velocity divergence in a regime of low $\sigma$. It is then of crucial interest to be able to derive the smoothing effects for the whole hierarchy of the cumulants. Other methods using various kinds of approximations have been used to derive the shape of this distribution function. Kofman (1991) and Kofman et al. (1993) use the Zel'dovich approximation (Zel'dovich 1970) for the large–scale dynamics to derive the shape of this distribution function but without taking into account the smoothing effects. The use of the Zel'dovich approximation, however, leads to wrong limits for the cumulants of the distribution as shown by Bernardeau & Kofman (1994) which undermines this method for a practical use.

In this paper, I report how it is possible to obtain the full hierarchy of the cumulants when the smoothing effects (for a top–hat window function only) are taken into account. Part 2 is devoted to a recall of the results that have been obtained when the smoothing effects are neglected. In Part 3, I present the general method to get the smoothing corrections for the density field. The resulting shape of the one–point probability distribution function of the density is examined in detail. Part 4 is devoted to similar calculations for the divergence of the velocity field.

## 2. The correlation hierarchy when the smoothing effects are neglected

Most of the results I am going to present in this part have been derived in detail in a previous paper (Bernardeau 1992). The notation, however, will be slightly updated.

As mentioned in the introduction, I assume that the large–scale structures of the Universe form by gravitational instabilities from Gaussian initial fluctuations and that the content of the universe behaves like a pressureless fluid. No assumptions are made about the cosmological parameters $\Omega$ and $\Lambda$. I will also assume that the multi streaming regime that may appear at small scales does not change the large–scale mass distribution and motion. The motion equations are then supposed to be given by the single stream approximation, that is that, at any location in space one can associate a unique peculiar velocity.

In the following the matter density field is described by the overdensity $\delta(\mathbf{x}) = \rho(\mathbf{x})/\overline{\rho} - 1$ and the peculiar velocity field, $\mathbf{u}(\mathbf{x})$, is, according to a general result of perturbation theory (Peebles 1980), assumed to be irrotational and then can be described by its divergence,

$$\theta(\mathbf{x}) = \frac{1}{H_0}\nabla.\mathbf{u}(\mathbf{x}). \tag{1}$$

### *2.1. Definition of the cumulants*

The basic objects of my analysis are the cumulants of the distribution function of $\delta$ and $\theta$. The $p^{th}$ cumulant of the distribution function of $\delta$, $\big\langle\delta^p\big\rangle_c$ is defined recursively from the $p^{th}$ moment following the rules,

$$\begin{aligned}
&\big\langle\delta\big\rangle_c = 0,\\
&\big\langle\delta^2\big\rangle_c = \big\langle\delta^2\big\rangle \equiv \sigma^2,\\
&\big\langle\delta^3\big\rangle_c = \big\langle\delta^3\big\rangle,\\
&\big\langle\delta^4\big\rangle_c = \big\langle\delta^4\big\rangle - 3\big\langle\delta^2\big\rangle_c^2,\\
&\big\langle\delta^5\big\rangle_c = \big\langle\delta^5\big\rangle - 10\big\langle\delta^2\big\rangle_c\big\langle\delta^3\big\rangle_c,\\
&\dots
\end{aligned}$$

and, in general, to obtain the $p^{th}$ cumulant one should consider all the decompositions of a set of $p$ points in its subsets (but the one being only the set itself); multiply, for each decomposition, the cumulants corresponding to each subset and subtract the results of all these products thus obtained from the $p^{th}$ moment. It is worth noticing that $\rho/\overline{\rho}$ and $\delta$ have the same cumulants. A similar definition applies for the cumulants of the distribution function of $\theta$. The second moment, $\sigma^2$, of the density distribution is related to the power spectrum of the fluctuations, $P(k)$. In the following I assume that the rms fluctuation is small enough to be accurately given by the linear theory so that,

$$\sigma^2(t) = \int \frac{\mathrm{d}^3\mathbf{k}}{(2\pi)^3} P(k) \left(\frac{D(t)}{D(t_0)}\right)^2, \tag{2}$$

where $D(t)$ is the time dependence of the growing mode and $t_0$ the present time if $P(k)$ is normalized to the present day large–scale fluctuations. The rms fluctuation of the divergence field, $\sigma_\theta$, is related to $\sigma$ through the function,

$$f(\Omega,\Lambda) = \frac{a}{D}\frac{\mathrm{d}D}{\mathrm{d}a}, \tag{3}$$

($a$ is the expansion factor) with

$$\sigma_\theta = f(\Omega,\Lambda)\sigma. \tag{4}$$

The function $f$ can be accurately approximated by $f(\Omega,\Lambda) = \Omega^{0.6}$ according to Peebles (1980) and Lahav et al. (1991).

Each cumulant then contains a new piece of information on the statistical properties of the fields related to the shape of the distribution function. A Gaussian distribution, for instance, is characterized by only one non–zero cumulant, its variance (see Bernardeau & Kofman 1994 for a more detailed discussion on the relationship between the shape of the density probability distribution function and the cumulants in the mildly nonlinear regime).

*2.2. Large–scale behavior of the cumulants*

The starting point of this analysis are the motion equations that describe the behavior of a pressureless fluid in the single stream approximation, the continuity, Euler and Poisson equations. The density field $\delta(\mathbf{x})$ can be expanded with respect to the initial fluctuation field, $\delta(\mathbf{x}) = \sum_{p=1}^{\infty}\delta^{(p)}(\mathbf{x})$ where $\delta^{(p)}$ is of order $p$ with respect to the initial fluctuation field. Using the motion equation it is possible to determine $\delta^{(p)}$ order by order. For an Einstein-de Sitter universe the solution can be expressed in terms of the Fourier transforms of the initial fluctuations, $\delta(\mathbf{k})$, so that (Goroff et al. 1986)

$$\begin{aligned}
\delta^{(1)}(\mathbf{x}) &= \int \frac{\mathrm{d}^3\mathbf{k}}{(2\pi)^{3/2}}\delta(\mathbf{k})\exp(\mathrm{i}\mathbf{k}\mathbf{x}) \\
&\dots \\
\delta^{(p)}(\mathbf{x}) &= \int \frac{\mathrm{d}^3\mathbf{k}_1}{(2\pi)^{3/2}}\dots\frac{\mathrm{d}^3\mathbf{k}_p}{(2\pi)^{3/2}}\,\delta(\mathbf{k}_1)\dots\delta(\mathbf{k}_p)\,N^{(p)}(\mathbf{k}_1,\dots,\mathbf{k}_p)\ \exp\left[\mathrm{i}(\mathbf{k}_1+\dots+\mathbf{k}_p)\mathbf{x}\right]
\end{aligned} \tag{5}$$

where $N^{(p)}(\mathbf{k}_1,\dots,\mathbf{k}_p)$ is an homogeneous function of the wave vectors $\mathbf{k}_1,\dots,\mathbf{k}_p$. Then, the *leading* term of the cumulant $\left\langle\delta^p\right\rangle_c$ when $\sigma$ is small is given by

$$\left\langle\delta^p\right\rangle_c = \sum_{\text{combinations},\ p(i)} \left\langle\prod_{i=1}^{p}\delta^{(p[i])}\right\rangle_c \tag{6}$$

where the sum is made over all the possible combinations $p(i)$, $i=1\dots p$, for which

$$p(i)\geq 1,\quad \sum_{i=1}^{p} p(i) = 2p-2. \tag{7}$$

This particular result is due to the hypothesis of Gaussian initial conditions. It is then possible to show that the terms in (6) are all products of the rms density fluctuation at the power $2p-2$ by some combination of the vertices,

$$\nu_p(a) \equiv \left\langle\delta^{(p)}\left(\delta^{(1)}\right)^p\right\rangle_c/\sigma^{2p} = \frac{p!}{(4\pi)^p}\int \mathrm{d}\Omega_1\dots\mathrm{d}\Omega_p\,N^{(p)}(\mathbf{k}_1,\dots,\mathbf{k}_p) \tag{8}$$

that is the angular average of the function $N^{(p)}(\mathbf{k}_1,\dots,\mathbf{k}_p)$. This angular average leads to keep only the monopole term of $N^{(p)}$. The coefficients $\nu_p$ are in general slightly time dependent but it just reflects a small change of these coefficients with the cosmological parameters (as shown in the following), and this dependence vanishes for an Einstein–de Sitter Universe. Then, each cumulant, $\left\langle\delta^p\right\rangle_c$, decreases with the variance like $\left\langle\delta^2\right\rangle^{p-1}$ so that the ratios $\left\langle\delta^p\right\rangle_c/\left\langle\delta^2\right\rangle^{p-1}$ have a finite limit,

$$\frac{\left\langle\delta^p\right\rangle_c}{\left\langle\delta^2\right\rangle^{p-1}} \rightarrow S_p(a) \quad \text{when} \quad \left\langle\delta^2\right\rangle \rightarrow 0. \tag{9}$$

The coefficients $S_p$ are closely related to the vertices $\nu_p$. Technically it corresponds to a tree summation (see Fry 1984, Bernardeau 1992). It is possible to get a closed analytical relationship between the vertices and the cumulants at the level of their generating functions. Let me define

$$\varphi(a,y) = \sum_{p=2}^{\infty} S_p(a) \frac{(-1)^{p-1}}{p!} y^p, \quad (S_2 = 1), \tag{10}$$

and

$$\mathcal{G}_\delta(a,\tau) = \sum_{p=1}^{\infty} \nu_p(a) \frac{(-\tau)^p}{p!}. \tag{11}$$

Then we have

$$\varphi(a,y) = y\mathcal{G}_\delta(a,\tau) - \frac{1}{2} y\tau \frac{\mathrm{d}}{\mathrm{d}\tau}\mathcal{G}_\delta(a,\tau), \quad \tau = -y\frac{\mathrm{d}}{\mathrm{d}\tau}\mathcal{G}_\delta(a,\tau), \tag{12}$$

These rules for tree summations have been given by De dominicis and Martin (1964) (see also, Des Cloiseaux and Jannink 1987) and has been used in an astrophysical context by Schaeffer (1985), by Bernardeau & Schaeffer (1992) and Bernardeau (1992) for the cumulants of the density distribution function.

The last step of the derivation of the $S_p$ parameters is the derivation of the $\mathcal{G}_\delta(a,\tau)$ function from the motion equations. Taking advantages of the geometrical averages in (8) it is possible to get a simple $2^{\text{nd}}$ order differential equation for $\mathcal{G}_\delta(a,\tau)$ that corresponds to the spherical collapse dynamics, *i.e.* , $\mathcal{G}_\delta$ is the density contrast of an object of linear overdensity $-\tau$. For an Einstein-de Sitter Universe the time dependence of the generating functions vanishes and the function $\mathcal{G}_\delta(\tau)$ is given by, when $\tau < 0$,

$$\mathcal{G}_\delta = \frac{9}{2}\frac{(\theta - \sin\theta)^2}{(1-\cos\theta)^3} - 1, \quad \tau = -\frac{3}{5}\left[\frac{3}{4}(\theta - \sin\theta)\right]^{2/3}, \tag{13$a$}$$

and when $\tau > 0$,

$$\mathcal{G}_\delta = \frac{9}{2}\frac{(\sinh\theta - \theta)^2}{(\cosh\theta - 1)^3} - 1, \quad \tau = \frac{3}{5}\left[\frac{3}{4}(\sinh\theta - \theta)\right]^{2/3}. \tag{13$b$}$$

For $\Omega \neq 1$ and $\Lambda = 0$, the analytical solutions are given in Bernardeau (1992, eq. [35]). In general ($\Lambda \neq 0$) there is no analytical solution for the spherical collapse dynamics. However, whatever the cosmological parameters, the function $\mathcal{G}_\delta(\tau)$ is well described by the function,

$$\mathcal{G}_\delta(\tau) = \left(1 + \frac{2\tau}{3}\right)^{-3/2} - 1, \tag{14}$$

showing that the time dependence of $\mathcal{G}_\delta(a,\tau)$ (and consequently the one of $S_p(a)$) is extremely weak. In the following I will widely use the approximation (14). One may wonder why such an approximation is good since it even fails to reproduce accurately the position of the singularity. Actually in the system (12) the values of interest for $\tau$ are $\tau > -0.42$ for which the approximation is extremely good. This limit in $\tau$ comes from the implicit equation which has a solution only in this domain, so that the vicinity of the singular point in (14) is not relevant.

Similar properties can also be derived for the divergence of the velocity field. Thus the large–scale limit of the cumulants of the velocity divergence probability distribution function follows the rules

$$\frac{\langle\theta^p\rangle_c}{\langle\theta^2\rangle^{p-1}} \longrightarrow T_p(a) \ \text{ when } \ \langle\theta^2\rangle \longrightarrow 0, \tag{15}$$

and the generating function of $T_p(a)$, $\varphi_\theta(a,y) = \sum_{p=2}^{\infty} T_p(a) y^p (-1)^{p-1}/p!$, is, as well, related to a generating function of vertices,

$$\varphi_\theta(a,y) = y\mathcal{G}_\theta(a,\tau) - \frac{1}{2}y\tau\frac{\mathrm{d}}{\mathrm{d}\tau}\mathcal{G}_\theta(a,\tau), \quad \tau = -y\frac{\mathrm{d}}{\mathrm{d}\tau}\mathcal{G}_\theta(a,\tau), \tag{16a}$$

with, in general,

$$\mathcal{G}_\theta[a, f(\Omega,\Lambda)\tau] = -\left[\frac{a\mathrm{d}}{\mathrm{d}a}\mathcal{G}_\delta(a,\tau) + f(\Omega,\Lambda)\frac{\tau\mathrm{d}}{\mathrm{d}\tau}\mathcal{G}_\delta(a,\tau)\right] / \left[1 + \mathcal{G}_\delta(a,\tau)\right]. \tag{16b}$$

As the time dependence of $\mathcal{G}_\delta(a,\tau)$ is weak we end up with,

$$\mathcal{G}_\theta[a, f(\Omega,\Lambda)\tau] = -f(\Omega,\Lambda)\frac{\tau\mathrm{d}}{\mathrm{d}\tau}\mathcal{G}_\delta(a,\tau)/\left[1 + \mathcal{G}_\delta(a,\tau)\right].$$

Using the approximative form of $f(\Omega,\Lambda)$, we obtain

$$\mathcal{G}_\theta(a,\tau) = \tau\left(1 + \frac{2\tau}{3\Omega^{0.6}}\right)^{-1}. \tag{17}$$

2.3. *The one–point probability distribution function*

Up to this point what have been obtained is an exact mathematical result for the large–scale limits of the cumulants. To construct the probability distribution function, I will now assume that the ratios $\langle\delta^p\rangle_c/\langle\delta^2\rangle^{p-1}$ actually equal their limit for the scales of interest. This approximation is expected to be valid as long as $\sigma$ is small compared to unity since the first corrective terms for the ratios would be of the order of $\sigma^2$ in a perturbative calculation.

The construction of the one–point probability distribution function of the density or of the divergence from the generating function of the cumulants has been discussed at length and for various regimes by Balian & Schaeffer (1989). The general result is that,

$$p(\delta)\mathrm{d}\delta = \int_{-\mathrm{i}\infty}^{+\mathrm{i}\infty} \frac{\mathrm{d}y}{2\pi\mathrm{i}\sigma^2} \exp\left[-\frac{\varphi(y)}{\sigma^2} + \frac{y\delta}{\sigma^2}\right]\mathrm{d}\delta, \tag{18}$$

where the integration is made in the complex plane. The limit of interest for this analysis is for $\sigma \longrightarrow 0$. In such a case and provided $\delta < \delta_c$ (Bernardeau 1992, Eq. 46),

$$p(\delta)\mathrm{d}\delta = \frac{1}{-\mathcal{G}_\delta'(\tau)}\left[\frac{1 - \tau\mathcal{G}_\delta''(\tau)/\mathcal{G}_\delta'(\tau)}{2\pi\sigma^2}\right]^{1/2} \exp\left(-\frac{\tau^2}{2\sigma^2}\right)\mathrm{d}\delta, \quad \mathcal{G}_\delta(\tau) = \delta\,, \tag{19}$$

where $\delta_c \approx 0.66$ is the value of the density contrast that cancels $1 - \tau\mathcal{G}_\delta''(\tau)/\mathcal{G}_\delta'(\tau)$. The function $\varphi(y)$ can as well be reobtained from the shape of the density probability distribution function,

$$\exp\left[-\frac{\varphi(y)}{\sigma^2}\right] = \int_{-1}^{\infty} \mathrm{d}\delta\, p(\delta)\exp\left[-\frac{y\delta}{\sigma^2}\right]. \tag{20}$$

The use of the form (18) for $p(\delta)$ allows indeed to revover $\varphi(y)$. But the approximate form (19), with the use of a second saddle point approximation, allows as well to recover the form (12) of $\varphi(y)$, showing that the form (19) still contains all the information on the generating function of the cumulants. This property will be of interest for the calculations in the next section.

A similar derivation can be done for the shape of the one–point distribution function of the velocity divergence.

## 3. The smoothing effects for the density field

### *3.1. The case of the skewness*

To illustrate the nature of the problem I will consider in this paragraph the case of the skewness of the density distribution function. It has been studied in detail in various papers cited in the introduction.

#### *3.1.1. Calculation in Lagragian space*

Let me first start with a Lagrangian description of the dynamics. In such a case the local density is obtained by the inverse of the Jacobian of the transformation, $J(\mathbf{q}) = |\partial\mathbf{x}/\partial\mathbf{q}|$, between the positions of the particles in Lagrangian space, $\mathbf{q}$, and their positions in real space, $\mathbf{x}$. The derivation of the third moment requires the calculation of the density field at the second order in the initial fluctuation field. The first order of the density fluctuation is simply given by $-J^{(1)}(\mathbf{q})$ and at the second order by $\delta^{(2)} = \left(J^{(1)}\right)^2 - J^{(2)}$. We then have (Bouchet et al. 1992),

$$\delta^{(2)}(\mathbf{q}) = \frac{5}{7}\left[\delta^{(1)}(\mathbf{q})\right]^2 + \frac{2}{7}\sum_{\alpha,\beta}\phi^{(1)}_{,\alpha\beta}(\mathbf{q})\phi^{(1)}_{,\alpha\beta}(\mathbf{q}), \tag{21}$$

where $\phi(\mathbf{q})$ if the gravitational potential at the Lagrangian position $\mathbf{q}$ and the subscript $_{,\alpha\beta}$ means the spatial derivatives with respect to the components $\alpha$ and $\beta$ of $\mathbf{q}$. It is more convenient to write this expression in Fourier space. Let me denote $\delta(\mathbf{k})$ the Fourier components of the initial density fluctuations (normalized to the present time), then

$$\begin{aligned}
\delta^{(1)}(\mathbf{q}) &= \int \frac{\mathrm{d}^3\mathbf{k}}{(2\pi)^{3/2}}\delta(\mathbf{k})\exp\left(\mathrm{i}\mathbf{q}.\mathbf{k}\right) \\
\delta^{(2)}(\mathbf{q}) &= \int \frac{\mathrm{d}^3\mathbf{k}_1}{(2\pi)^{3/2}}\frac{\mathrm{d}^3\mathbf{k}_2}{(2\pi)^{3/2}}\delta(\mathbf{k}_1)\delta(\mathbf{k}_2)\left[\frac{5}{7}+\frac{2}{7}\left(\frac{\mathbf{k}_1.\mathbf{k}_2}{k_1k_2}\right)^2\right]\exp\left[\mathrm{i}\mathbf{q}.(\mathbf{k}_1+\mathbf{k}_2)\right].
\end{aligned} \tag{22}$$

The third moment of the density distribution is then given by $3\big\langle\left(\delta^{(1)}\right)^2\delta^{(2)}\big\rangle$ (which is its dominant term in perturbation theory). The calculation of this ensemble average reads (once the symmetry factors have been taken into account)

$$\big\langle\delta^3\big\rangle = 6\int\frac{\mathrm{d}^3\mathbf{k}_1}{(2\pi)^3}\frac{\mathrm{d}^3\mathbf{k}_2}{(2\pi)^3}\,P(k_1)P(k_2)\left[\frac{5}{7}+\frac{2}{7}\left(\frac{\mathbf{k}_1.\mathbf{k}_2}{k_1k_2}\right)^2\right]. \tag{23}$$

The integral over the angle between $\mathbf{k}_1$ and $\mathbf{k}_2$ in (23) is simple in the absence of window function and leads to the result,

$$\big\langle\delta^3\big\rangle = \frac{34}{7}\big\langle\delta^2\big\rangle^2. \tag{24}$$

#### *3.1.1. Filtering in Lagragian space*

Let me then consider the skewness of the density distribution function for a filtering at a given mass scale. The density is simply defined as the ratio of the initial volume by the final volume occupied by a certain amount of matter. I will only consider the case of a top–hat filter that means that I will examine the statistics the inverse of the volume of particles that lie, in Lagrangian space, in a sphere of given radius $R_0$. The volume occupied by these particles all along the evolution is given by the integral of the Jacobian over the sphere of radius $R_0$. The top–hat filter should then be applied to the field $J(\mathbf{q})$ and not to the field $\delta(\mathbf{q})$. Once it has been done the filtered density field at first and second order read

$$
\begin{aligned}
\delta^{(1)}_{R_0}(\mathbf{q}) &= \int \frac{\mathrm{d}^3\mathbf{k}}{(2\pi)^{3/2}}\delta(\mathbf{k})\exp\left(\mathrm{i}\mathbf{q}.\mathbf{k}\right) W_{\mathrm{TH}}(k\,R_0)\\
\delta^{(2)}_{R_0}(\mathbf{q}) &= \int \frac{\mathrm{d}^3\mathbf{k}_1}{(2\pi)^{3/2}}\frac{\mathrm{d}^3\mathbf{k}_2}{(2\pi)^{3/2}}\delta(\mathbf{k}_1)\delta(\mathbf{k}_2)\exp\left[\mathrm{i}\mathbf{q}.(\mathbf{k}_1+\mathbf{k}_2)\right]\\
&\times\left(W_{\mathrm{TH}}(k_1\,R_0)W_{\mathrm{TH}}(k_2\,R_0)-\frac{2}{7}\left[1-\left(\frac{\mathbf{k}_1.\mathbf{k}_2}{k_1k_2}\right)^2\right]W_{\mathrm{TH}}\left(\vert\mathbf{k}_1+\mathbf{k}_2\vert\;R_0\right)\right),
\end{aligned}
\tag{25}
$$

where $W_{\mathrm{TH}}(k\,R_0)$ is the Fourier transform of a top–hat window function,

$$
W_{\mathrm{TH}}(k\,R_0) = \frac{3}{(kR_0)^3}\left[\sin(kR_0)-kR_0\cos(kR_0)\right]. \tag{26}
$$

The integral over the wave vectors $\mathbf{k}_1$ and $\mathbf{k}_2$ is dramatically changed in,

$$
\begin{aligned}
\left\langle\delta^3_{R_0}\right\rangle = 6\int &\frac{\mathrm{d}^3\mathbf{k}_1}{(2\pi)^3}\frac{\mathrm{d}^3\mathbf{k}_2}{(2\pi)^3}\,P(k_1)P(k_2)W_{\mathrm{TH}}(k_1\,R_0)W_{\mathrm{TH}}(k_2\,R_0)\\
&\left(W_{\mathrm{TH}}(k_1\,R_0)W_{\mathrm{TH}}(k_2\,R_0)-\frac{2}{7}W_{\mathrm{TH}}\left(\vert\mathbf{k}_1+\mathbf{k}_2\vert\;R_0\right)\left[1-\left(\frac{\mathbf{k}_1.\mathbf{k}_2}{k_1k_2}\right)^2\right]\right).
\end{aligned}
\tag{27}
$$

A geometrical property of the top–hat window function given by Bernardeau (1994b, Eq. [A.5]), however, shows that the integral over the angle between the wave vectors is also simple and leads exactly to the same coefficient as previously,

$$
\left\langle\delta^3_{R_0}\right\rangle = \frac{34}{7}\left\langle\delta^2_{R_0}\right\rangle^2,
$$

showing that the filtering at a given *mass* scale does not change the skewness.

### *3.1.3. Filtering in Eulerian space*

In practice, however, we are more interested in the behavior of the distribution function in Eulerian space, for a given smoothing radius, $R_0$. When the density field is expressed in term of the Eulerian coordinates, $\mathbf{x}$, rather than the Lagrangian coordinates, its first and second order read,

$$
\begin{aligned}
\delta^{(1)}(\mathbf{x}) &= \int \frac{\mathrm{d}^3\mathbf{k}}{(2\pi)^{3/2}}\delta(\mathbf{k})\exp\left(\mathrm{i}\mathbf{x}.\mathbf{k}\right)\\
\delta^{(2)}(\mathbf{x}) &= \int \frac{\mathrm{d}^3\mathbf{k}_1}{(2\pi)^{3/2}}\frac{\mathrm{d}^3\mathbf{k}_2}{(2\pi)^{3/2}}\delta(\mathbf{k}_1)\delta(\mathbf{k}_2)\exp\left[\mathrm{i}\mathbf{x}.(\mathbf{k}_1+\mathbf{k}_2)\right]\left[\frac{5}{7}-\frac{\mathbf{k}_1.\mathbf{k}_2}{k_1^2}+\frac{2}{7}\left(\frac{\mathbf{k}_1.\mathbf{k}_2}{k_1k_2}\right)^2\right].
\end{aligned}
\tag{28}
$$

The new term in $\delta^{(2)}$ comes from the expression of $\delta^{(1)}(\mathbf{q})$ and is due to the change of variable from $\mathbf{q}$ to $\mathbf{x}$. If one calculates the skewness from the expression (28), that is without smoothing, one gets the coefficient 34/7 as in equation (27) for the Lagrangian case. The smoothing of the density field changes, however, the result. Indeed, after filtering, the third moment for the density field is given by,

$$\begin{aligned}\langle\delta_{R_0}^3\rangle = 6\int \frac{\mathrm{d}^3\mathbf{k}_1}{(2\pi)^3}\frac{\mathrm{d}^3\mathbf{k}_2}{(2\pi)^3}\,P(k_1)P(k_2)W_{\rm TH}(k_1\,R_0)W_{\rm TH}(k_2\,R_0)\\ \times\,W_{\rm TH}\left(\vert\mathbf{k}_1+\mathbf{k}_2\vert\,R_0\right)\left[\frac{5}{7}-\frac{\mathbf{k}_1.\mathbf{k}_2}{k_1^2}+\frac{2}{7}\left(\frac{\mathbf{k}_1.\mathbf{k}_2}{k_1k_2}\right)^2\right].\end{aligned}\tag{29}$$

This expression can then be calculated using a second property of the top–hat window function given by Bernardeau (1994b, Eq. [A.6]). It leads to

$$\langle\delta_{R_0}^3\rangle=\left(\frac{34}{7}+\gamma_1\right)\langle\delta_{R_0}^2\rangle^2,\tag{30a}$$

with

$$\gamma_1=\frac{\mathrm{d}\log\sigma^2(R_0)}{\mathrm{d}\log R_0}.\tag{30b}$$

There is here a corrective term that depends on the scale dependence of rms fluctuation. The origin of this correction can now be understood as a shift in mass scale. Indeed, for a given mass scale the skewness remains uncorrected as the analysis in the Lagrangian space demonstrated it. However, the filtering process in Eulerian space mixes different mass scales: a given density $\rho$ measured at a radius $R_0$ corresponds to a mass scale of $\rho\ 4\pi R_0^3/3$ (and not $4\pi R_0^3/3$). When the rms fluctuation decreases with mass, less high density regions and more low density regions can be expected than if $\sigma$ were scale independent. The skewness being sensitive to the asymmetry between the large density contributions and the low density contributions is then expected to decrease. The result (30) quantifies this departure.

### *3.2. The correlation hierarchy with the smoothing effects*

#### *3.2.1 Filtering in Lagrangian space*

Once again I will start by the density distribution in Lagrangian space, that is for a given mass scale. In a Lagrangian description the density is given by the inverse of the Jacobian. One can write any order $p$ of the expansion of the Jacobian with the initial displacement field,

$$J^{(p)}(t)=\int\frac{\mathrm{d}^3\mathbf{k}_1}{(2\pi)^{3/2}}\dots\frac{\mathrm{d}^3\mathbf{k}_p}{(2\pi)^{3/2}}\delta_{\mathbf{k}_1}\dots\delta_{\mathbf{k}_p}\ j^{(p)}(\mathbf{k}_1,\dots,\mathbf{k}_p)\ \exp\left[\mathrm{i}\mathbf{x}(\mathbf{k}_1+\dots+\mathbf{k}_p)\right]$$

where $j^{(p)}(\mathbf{k}_1,\dots,\mathbf{k}_p)$ is a function of the angular parts of the wave vectors $\mathbf{k}_1,\dots,\mathbf{k}_p$ only. For instance for $p=2$ we have $j^{(2)}=2/7(1-u_{12}^2)$ with $u_{12}=\mathbf{k}_1.\mathbf{k}_2/(k_1\,k_2)$. To get the expression of $J_{R_0}^{(p)}$ when the field is smoothed at the radius $R_0$ one has to multiply the function $j^{(p)}(\mathbf{k}_1,\dots,\mathbf{k}_p)$ by the Fourier transform of the window function applied to the sum of the wave vectors, $W_{\rm TH}\left[\vert\mathbf{k}_1+\dots+\mathbf{k}_p\vert\ R_0\right]$. The result obtained in the appendix C of Bernardeau (1994a) is a geometrical property of the function $j^{(p)}(\mathbf{k}_1,\dots,\mathbf{k}_p)$. Indeed the integral over the angular parts of the wave vectors gives

$$\begin{aligned}\int\mathrm{d}\Omega_1\dots\mathrm{d}\Omega_p W_{\rm TH}\left[\vert\mathbf{k}_1+\dots+\mathbf{k}_p\vert\ R_0\right]j^{(p)}(\mathbf{k}_1,\dots,\mathbf{k}_p)=\\ (4\pi)^p\ \frac{j_p(t)}{p!}W_{\rm TH}(k_1\,R_0)\dots W_{\rm TH}(k_p\,R_0).\end{aligned}\tag{31}$$

where $j_p$ contains only the monopole term of $j^{(p)}(\mathbf{k}_1,\dots,\mathbf{k}_p)$, *i.e.* , the quantity you would have obtained simply by dropping the function $W_{\rm TH}$ in (27). The integral over the angles then lead to a factorization of the $k_1,\dots,k_p$ dependence and the result is exactly what would have been obtained if the filtering had been made on the initial field that is replacing $P(k)$ by $W^2(k)\,P(k)$. The continuity equation, together with the properties of the generating functions given by Bernardeau (1992, Eq. [26]), gives the generating function of the vertices for the density field,

$$\mathcal{G}_\delta(\tau)=\left(1+\sum_{p=1}^{\infty}j_p\frac{(-\tau)^p}{p!}\right)^{-1}-1 \tag{32}$$

which is exactly the function (11) obtained when the smoothing was neglected. So we show here that the filtering preserves the shape of the high–order correlations and does not change the generating function of the vertices. The property of the previous paragraph can then be generalized: the smoothing does not change the values of the $S_p$ parameters when it is done at a given mass scale.

### 3.2.2. Filtering in Eulerian space

The difficulty then lies in the derivation of the $S_p$ parameters for the smoothed density field in real space. From the analysis of the skewness it appears that it will be due to a mixing of different masses. In general the mapping between the Lagragian space and the Eulerian space is rather complicated. However, an exact relationship between the probability distribution functions of the filtered density field in Lagrangian and in Eulerian space is not necessary. What is needed is a trick to get the generating function of the large–scale cumulants that takes into account the mass mixing. Let me choose at random a point of matter $\mathbf{x}_0$ in the field. I can define $p_{Ec}(R_0,M)\mathrm{d}M$ the probability distribution function of the mass contained within the radius $R_0$ centered on $\mathbf{x}_0$. On the other hand, for a given mass $M_0$ it exists a unique radius $R$ so that the mass contained in a cell of radius $R$ centered on $\mathbf{x}_0$ is exactly $M_0$. Let $p_{Lc}(M_0,R)\mathrm{d}R$ be the probability distribution function of such a radius. Then the probability that a given volume $V_0$ of radius $R_0$, contains an amount of matter greater than $M_0$ is also the probability that the mass $M_0$ occupies a volume smaller than $V_0$. As a result we have

$$\int_{M_0}^{\infty}p_{Ec}(R_0,M)\;\mathrm{d}M=\int_0^{R_0}p_{Lc}(M_0,R)\;\mathrm{d}R. \tag{33a}$$

Both $p_{Ec}$ and $p_{Lc}$ can be seen as density probability distribution functions for respectively a fixed radius $R_0$ or a fixed mass $M_0$, so that the previous relation also reads

$$\int_{\delta_0}^{\infty}p_{Ec}(R_0,\delta)\mathrm{d}\delta=\int_{\delta_0}^{\infty}p_{Lc}(M_0,\delta)\mathrm{d}\delta, \tag{33b}$$

with $\delta_0=3M_0/(4\pi R_0\overline{\rho})-1$. For a continuous field the relationship between the function $p_{Ec}$ and the Eulerian smoothed density distribution function $p_E$ is simple and reads, $p_{Ec}(\delta)\mathrm{d}\delta=(1+\delta)\;p_E(\delta)\mathrm{d}\delta$. In general the relation between $p_{Lc}$ and the Lagrangian smoothed density distribution function $p_L$ is complicated. This is due to the fact that the particles that were initially in a spherical region end not necessarily in a spherical region in Eulerian space. However, as recalled in §2 the $S_p$ parameters are only determined by the spherical collapse dynamics, and as seen for the skewness the smoothing corrections come from the mass scale mixing. The possible aspherical movement of the matter is thus not expected to change the value of the large scale limit of the $S_p$ parameters. I will then neglect it and assume that $p_{Lc}=p_L$. Using the relation (33b) it is then possible to relate $p_E$ and $p_L$. By differentiating with respect to the threshold $\delta_0$ for a fixed radius $R_0$, we get

$$(1+\delta_0)p_E(\delta_0)=p_L(\delta_0)+\int_{\delta_0}^{\infty}\overline{\rho}\,V_0\frac{\mathrm{d}}{\mathrm{d}M_0}p_L(\delta)\mathrm{d}\delta.$$

The large–scale values of the $S_p$ parameters of $p_L(\delta)\mathrm{d}\delta$ are known and are the ones of the unsmoothed density field. We can then use the relation (18) to derive the expression of the density distribution function in Lagrangian space. We then obtain for the distribution function in Eulerian space,

$$\begin{aligned} p_E(\delta_0) = \frac{1}{1+\delta_0}\int_{-\mathrm{i}\infty}^{+\mathrm{i}\infty}\frac{\mathrm{d}y}{2\pi\mathrm{i}y}\left(-\frac{\varphi(y)}{\sigma^2(M_0)}+\frac{y\delta}{\sigma^2(M_0)}\right)\overline{\rho}\,V_0\frac{\mathrm{d}\sigma^2(M_0)}{\mathrm{d}M_0} \\ \times\exp\left[-\frac{\varphi(y)}{\sigma^2(M_0)}+\frac{y\delta}{\sigma^2(M_0)}\right]. \end{aligned} \tag{34}$$

Let me replace the mass dependence of $\sigma$ by a scale dependence, so that $3\overline{\rho}\,V_0\mathrm{d}\sigma^2(M_0)/\mathrm{d}M_0 = \mathrm{d}\sigma^2(R_0)/\mathrm{d}\log(R_0)$. The generating function of the cumulants for the smoothed density field at a given scale $R_0$ can then be obtained directly from the equation (34),

$$\exp\left[-\frac{\varphi_E(y)}{\sigma^2(R_0)}\right] = \int_{-1}^{\infty}\mathrm{d}\delta p_E(\delta)\exp\left[-\frac{\delta y}{\sigma^2(R_0)}\right]. \tag{35}$$

Actually we do not need to calculate exactly the expression of the right side of (35) but only the limit of its logarithm when $\sigma=0$. One can then use safely the saddle point approximation to integrate over $y$ to get $p_E(\delta)$ from (34). It leads to

$$\begin{aligned} \exp\left[-\frac{\varphi_E(y)}{\sigma^2(R_0)}\right] &= \frac{R_0}{3(1+\delta_0)}\frac{\mathrm{d}\sigma^2(R_0)}{\mathrm{d}R_0}\int_{-1}^{\infty}\frac{-\mathrm{d}\delta}{\mathcal{G}_\delta'(\tau)}\left[\delta+\frac{\tau}{\mathcal{G}_\delta'(\tau)}\delta-\frac{\tau^2}{2}\right] \\ &\times\left[\frac{1-\tau\mathcal{G}_\delta''(\tau)/\mathcal{G}_\delta'(\tau)}{2\pi\sigma^2([1+\delta]^{1/3}R_0)}\right]^{1/2}\exp\left[-\frac{\delta y}{\sigma^2(R_0)}-\frac{\tau^2}{2\sigma^2([1+\delta]^{1/3}R_0)}\right] \end{aligned} \tag{36}$$

$$\text{with}\quad \mathcal{G}_\delta(\tau)=\delta. \tag{37}$$

Let me then define the function $\mathcal{G}_\delta^{\mathcal{S}}$ by the implicit equation,

$$\mathcal{G}_\delta^{\mathcal{S}}(\tau^{\mathcal{S}}) = \mathcal{G}_\delta\left\{\tau^{\mathcal{S}}\ \frac{\sigma\left([1+\mathcal{G}_\delta^{\mathcal{S}}(\tau^{\mathcal{S}})]^{1/3}R_0\right)}{\sigma(R_0)}\right\}. \tag{38}$$

When $\tau_s = \tau\ \sigma(R_0)/\sigma\left([1+\mathcal{G}_\delta^{\mathcal{S}}(\tau^{\mathcal{S}})]^{1/3}R_0\right)$ the term under the exponential reads $[\delta y + (\tau^{\mathcal{S}})^2/2]/\sigma^2(R_0)$, $\mathcal{G}_\delta^{\mathcal{S}}(\tau^{\mathcal{S}})=\delta$. The integration over $\delta$ of (37) can be made by a second saddle point approximation (similar to the one used to obtain the expression (11) from the approximate form (19)). The saddle point is then given by,

$$\delta = \mathcal{G}_\delta^{\mathcal{S}}(\tau^{\mathcal{S}}), \tag{39}$$

where $\tau^{\mathcal{S}}$ is solution of the implicit equation,

$$y = -\tau^{\mathcal{S}}\frac{\mathrm{d}\tau^{\mathcal{S}}}{\mathrm{d}\delta} = \frac{-\tau^{\mathcal{S}}}{\frac{\mathrm{d}}{\mathrm{d}\tau}\mathcal{G}_\delta^{\mathcal{S}}(\tau^{\mathcal{S}})}. \tag{40}$$

Once the logarithm has been taken, the limit $\sigma=0$ conserves only the term that is under the exponential at the saddle point position so that $\varphi_E(y) = y\mathcal{G}_\delta^{\mathcal{S}}(\tau^{\mathcal{S}})+(\tau^{\mathcal{S}})^2/2$ and $\tau^{\mathcal{S}}$ is given by the saddle point position (40). As a result the series of the parameters $S_p$, when the smoothing corrections are taken into account, is still obtained by a generating function of vertices. The function $\mathcal{G}_\delta$ has simply been replaced by $\mathcal{G}_\delta^{\mathcal{S}}$ defined by the equation (38).

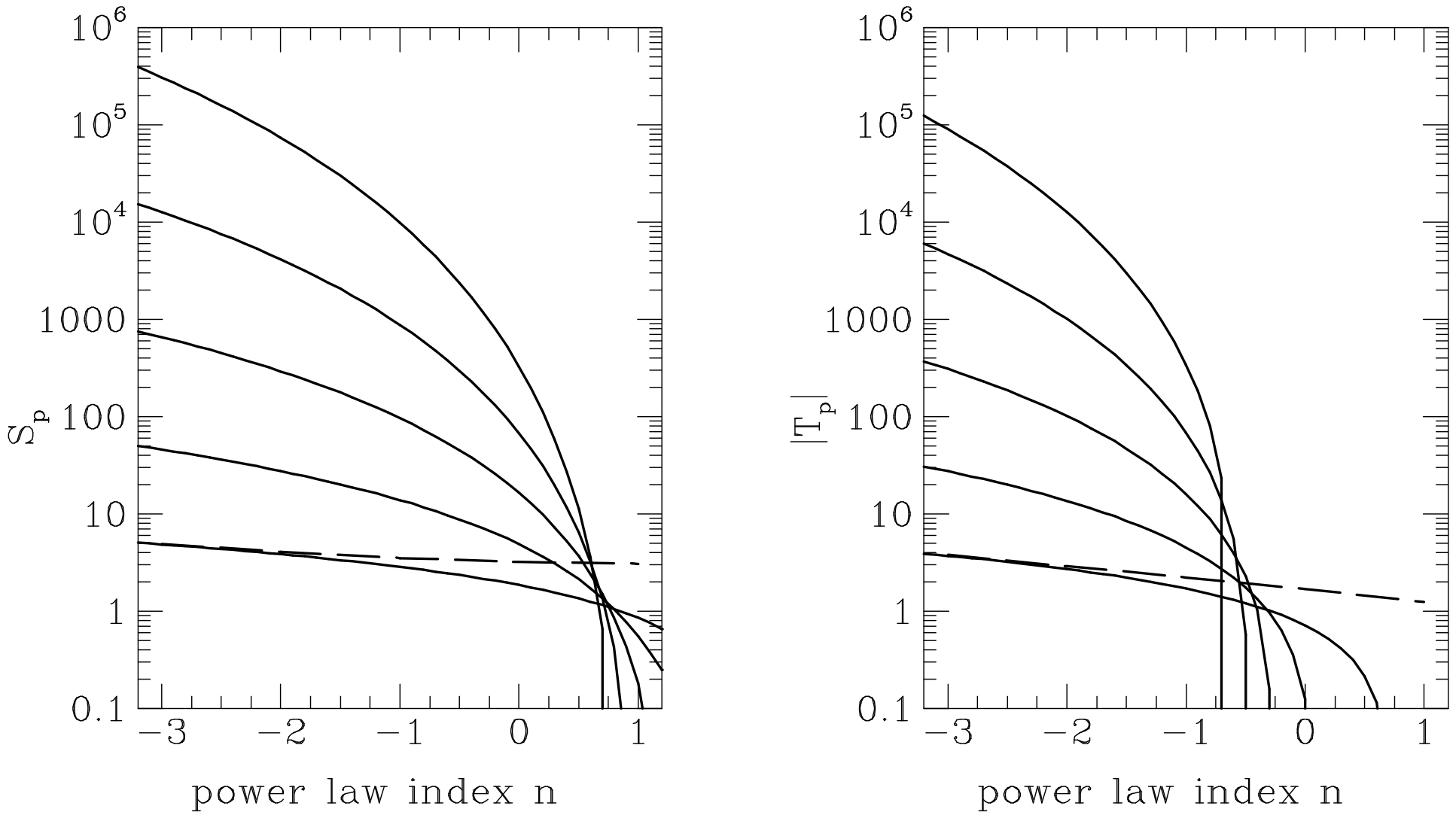

**Fig 1.** The coefficients $S_p$ (left panel) and $|T_p|$ (right panel) (Eqs. [9,15]) as a function of the power law index for $p = 3, \dots, 7$ (from bottom to top). The dashed lines correspond to the coefficients $S_3$ and $T_3$ obtained for a Gaussian window function by respectively Juszkiewicz et al. (1993a) and Bernardeau et al. (1994).

The generating function of the cumulants of the smoothed density field is thus given by the relation,

$$\varphi^{\mathcal{S}}(y) = y\mathcal{G}_\delta^{\mathcal{S}}(\tau) - \frac{1}{2} y\tau \frac{\mathrm{d}}{\mathrm{d}\tau}\mathcal{G}_\delta^{\mathcal{S}}(\tau), \quad \tau = -y\frac{\mathrm{d}}{\mathrm{d}\tau}\mathcal{G}_\delta^{\mathcal{S}}(\tau). \tag{41}$$

In the following I use the superscript $^{\mathcal{S}}$ for the quantities that have been obtained when the filtering effects are taken into account. The relation (41) is the central result of this paper. All the applications that are presented in the following derive from the equations (41), (38) and (13).

### *3.3. Example of cumulants*

By expanding the function $\varphi^{\mathcal{S}}(y)$ with respect to $y$ it is then possible to recover the full hierarchy of the cumulants. We have already noticed that they will be changed due to the variation of $\sigma$ with scale. If $\sigma$ were scale independent, which corresponds of a power law index $n = -3$, then the cumulants are identical to the ones found without smoothing. In this paragraph I will express the dependence of $\sigma$ with scale through its logarithmic derivatives with scale,

$$\gamma_p = \frac{\mathrm{d}^p \log \sigma^2(R_0)}{\mathrm{d}\log^p R_0}. \tag{42}$$

A tedious calculation (to be done with a symbolic calculator), can then give the expression of the first few $S_p$ parameters as a function of $\gamma_p$,

$$
\begin{aligned}
S_3 &= \frac{34}{7} + \gamma_1, \\
S_4 &= \frac{60712}{1323} + \frac{62\,\gamma_1}{3} + \frac{7\,{\gamma_1}^2}{3} + \frac{2\,\gamma_2}{3}, \\
S_5 &= \frac{200575880}{305613} + \frac{1847200\,\gamma_1}{3969} + \frac{6940\,{\gamma_1}^2}{63} + \frac{235\,{\gamma_1}^3}{27} + \frac{1490\,\gamma_2}{63} + \frac{50\,\gamma_1\,\gamma_2}{9} + \frac{10\,\gamma_3}{27}, \\
S_6 &= 12650 + 12330\,\gamma_1 + 4512\,{\gamma_1}^2 + 734.0\,{\gamma_1}^3 + 44.81\,{\gamma_1}^4 + 775.8\,\gamma_2 \\
&\quad + 375.9\,\gamma_1\,\gamma_2 + 45.56\,{\gamma_1}^2\,\gamma_2 + 3.889\,{\gamma_2}^2 + 20.05\,\gamma_3 + 4.815\,\gamma_1\,\gamma_3 + 0.1852\,\gamma_4, \\
S_7 &= 307810 + 383000\,\gamma_1 + 190700\,{\gamma_1}^2 + 47460\,{\gamma_1}^3 + 5914\,{\gamma_1}^4 + 294.8\,{\gamma_1}^5 \\
&\quad + 27340\,\gamma_2 + 20300\,\gamma_1\,\gamma_2 + 5026\,{\gamma_1}^2\,\gamma_2 + 414.8\,{\gamma_1}^3\,\gamma_2 + 358.1\,{\gamma_2}^2 \\
&\quad + 88.15\,\gamma_1\,{\gamma_2}^2 + 902.6\,\gamma_3 + 443.3\,\gamma_1\,\gamma_3 + 54.44\,{\gamma_1}^2\,\gamma_3 + 7.778\,\gamma_2\,\gamma_3 + 14.20\,\gamma_4 \\
&\quad + 3.457\,\gamma_1\,\gamma_4 + 0.08642\,\gamma_5, \\
&\ldots
\end{aligned}
$$

These results are valid for an Einstein–de Sitter Universe and have been obtained from the expression (13) from $\mathcal{G}_\delta(\tau)$.

The first two coefficients have already been given by Bernardeau (1994b) from a direct calculation using perturbation theory up to the third order. It confirms that the method presented here, despite its technical simplicity (and the lack of a rigorous derivation), can give straightforwardly the full series of the coefficients. These coefficients are presented as a function of $n$ assuming a power spectrum of index $n$ in Fig. 1. Note that when $n$ is close to 1 they vanish. In such a regime the results presented here should be used with caution: the high–order cumulants are expected to be dominated by higher order contributions of the perturbation theory.

### *3.4. The resulting shape of the density distribution function*

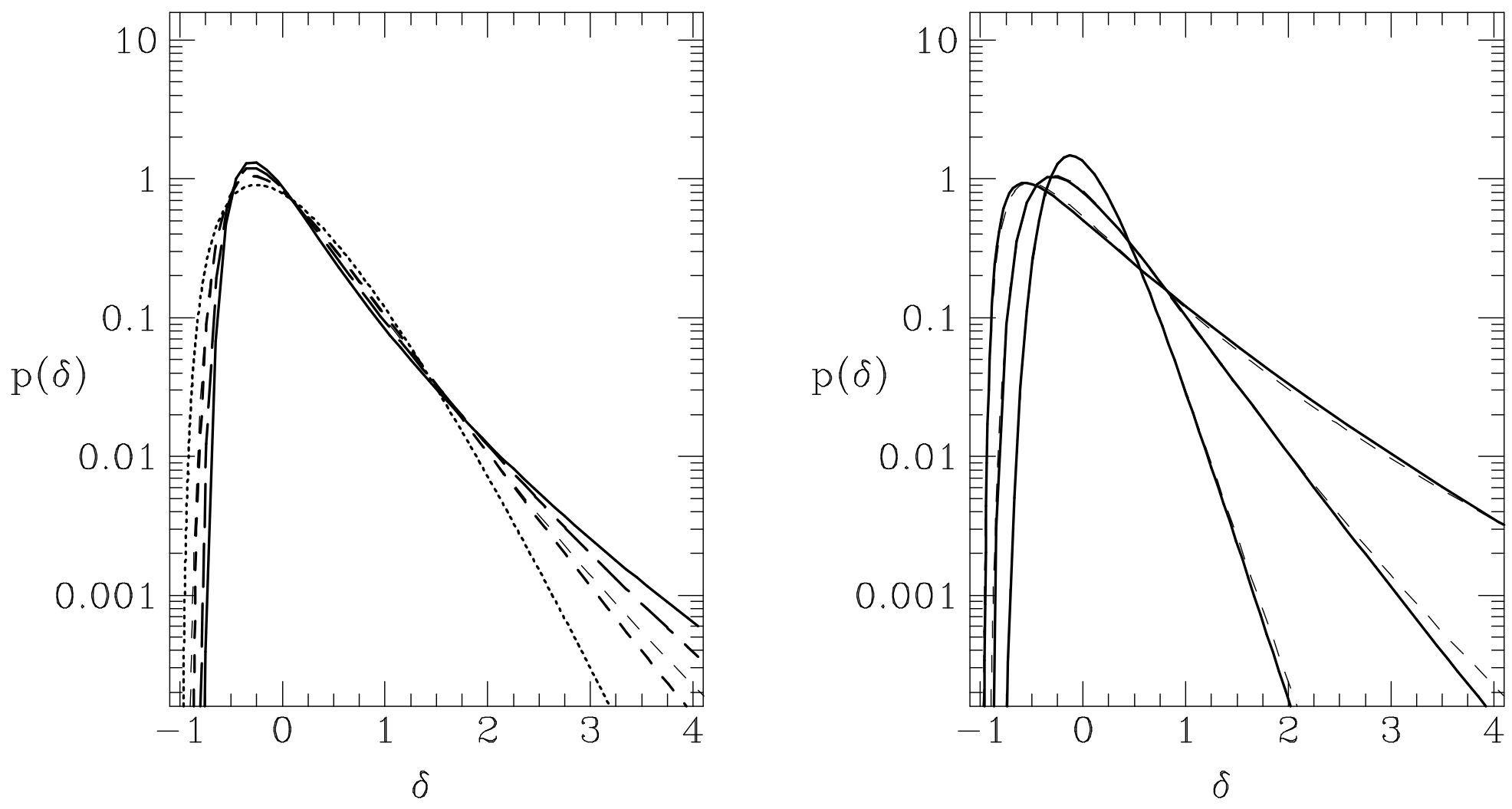

**Fig 2.** The one–point probability distribution function of the density contrast in the quasilinear regime. In the left panel $\sigma = 0.5$ and the power law index varies, $n$ =0 (dotted line), -1 (dashed line), -2 (long dashed line), -3 (solid line). In the right panel, $n = -1$ and $\sigma$ varies ($\sigma$ =0.3, 0.5, 0.8). In both cases the thin dashed lines correspond to the lognormal distribution (Eq. [52]).

Let me first make a remark on the theoretical expression of the density distribution function is Eulerian space. For the calculation of $p_E(\delta)$ from $p_L(\delta)$ in (34) it was assumed that the ratios $\left\langle\delta^p\right\rangle_c/\left\langle\delta^2\right\rangle^2$ are constant in Lagrangian space. This form, however, does not imply that the similar ratios obtained for the density distribution function in Eulerian space follow the same scaling rules, although their limits when $\sigma = 0$ are the exact ones. In the absence of any other physical arguments I will assume in the following that these ratios are actually constant in Eulerian space to derive the density probability distribution function. That means that the expression (18) (with $\varphi^{\mathcal{S}}(y)$ instead of $\varphi(y)$), rather than the expression (34), should be used to derive the shape of the density probability distribution function. Comparisons with numerical simulations presented in the next subsection prove that the hypothesis of constant ratios is extremely powerful.

For the sake of simplicity, I assume in the following that the scale dependence of the rms fluctuation can be approximated by a power law behavior, $\sigma(R_0) \propto R_0^{-(n+3)/2}$. In such a case the power spectrum would have a power law behavior,

$$P(k) \propto k^n. \tag{43}$$

This is a useful approximation which is accurate for most of the power spectrum of interest. (It is however possible to do the subsequent calculations without it.) In such a case, $\gamma_1 = -(n+3)$, $\gamma_p = 0$ when $p > 2$, and so

$$\mathcal{G}_\delta^{\mathcal{S}}(\tau) = \mathcal{G}_\delta\left(\tau\left[1+\mathcal{G}_\delta^{\mathcal{S}}(\tau)\right]^{-(n+3)/6}\right). \tag{44}$$

The use of the approximation (14) for the function $\mathcal{G}_\delta(\tau)$ leads to

$$\tau = \frac{3}{2}\left(1+\mathcal{G}_\delta^{\mathcal{S}}\right)^{(n+3)/6}\left[\left(1+\mathcal{G}_\delta^{\mathcal{S}}\right)^{-2/3}-1\right]. \tag{45}$$

Note that for $n = -1$ this relation can be inverted in $\mathcal{G}_\delta^{\mathcal{S}} = [-\tau/3 + (1+\tau^2/9)^{1/2}]^3 - 1$. Integrating the equation (18) one can obtain the shape of the probability distribution function for different values of $n$ and of $\sigma$. Some of these functions are given in Fig. 2.

It is also possible to find analytical approximations of the density probability distribution function. One key feature to get these forms is to determine whether the equation $\tau\mathcal{G}_\delta^{\mathcal{S}\,''}(\tau)/\mathcal{G}_\delta^{\mathcal{S}\,'}(\tau) = 1$ has a solution or not. As can be checked from equation (45) this equation has a solution only if $n < 0$. Note that this limit actually is not due to the approximation (14) but is exact for all cosmological parameters. In table 1, I present the value of $\tau$ that is solution of this equation and the values of the parameters of interest at this point. It corresponds indeed to a singular point in $y = y_s$ for the function $\varphi^{\mathcal{S}}(y)$ and close to this value $\varphi(y)$ behaves like,

$$\varphi^{\mathcal{S}}(y) - \varphi_s \simeq -a_s(y-y_s)^{3/2}. \tag{46}$$

The values of $y_s$, $a_s$ and $\varphi_s$ are given in table 1 for different values of $n$ and will be of interest in the following.

**Table 1.** *The parameters of the critical point as a function of the spectral index, n, for the density distribution.*

| $n$ | $\delta_c$ | $y_s$ | $a_s$ | $\varphi_s$ |
|---|---|---|---|---|
| $-3$ | .656 | $-0.184$ | 1.84 | $-0.030$ |
| $-2.5$ | 0.804 | $-0.213$ | 2.21 | $-0.041$ |
| $-2$ | 1.034 | $-0.253$ | 2.81 | $-0.058$ |
| $-1.5$ | 1.443 | $-0.310$ | 3.93 | $-0.093$ |
| $-1$ | 2.344 | $-0.401$ | 6.68 | $-0.172$ |
| $-0.5$ | 5.632 | $-0.574$ | 18.94 | $-0.434$ |

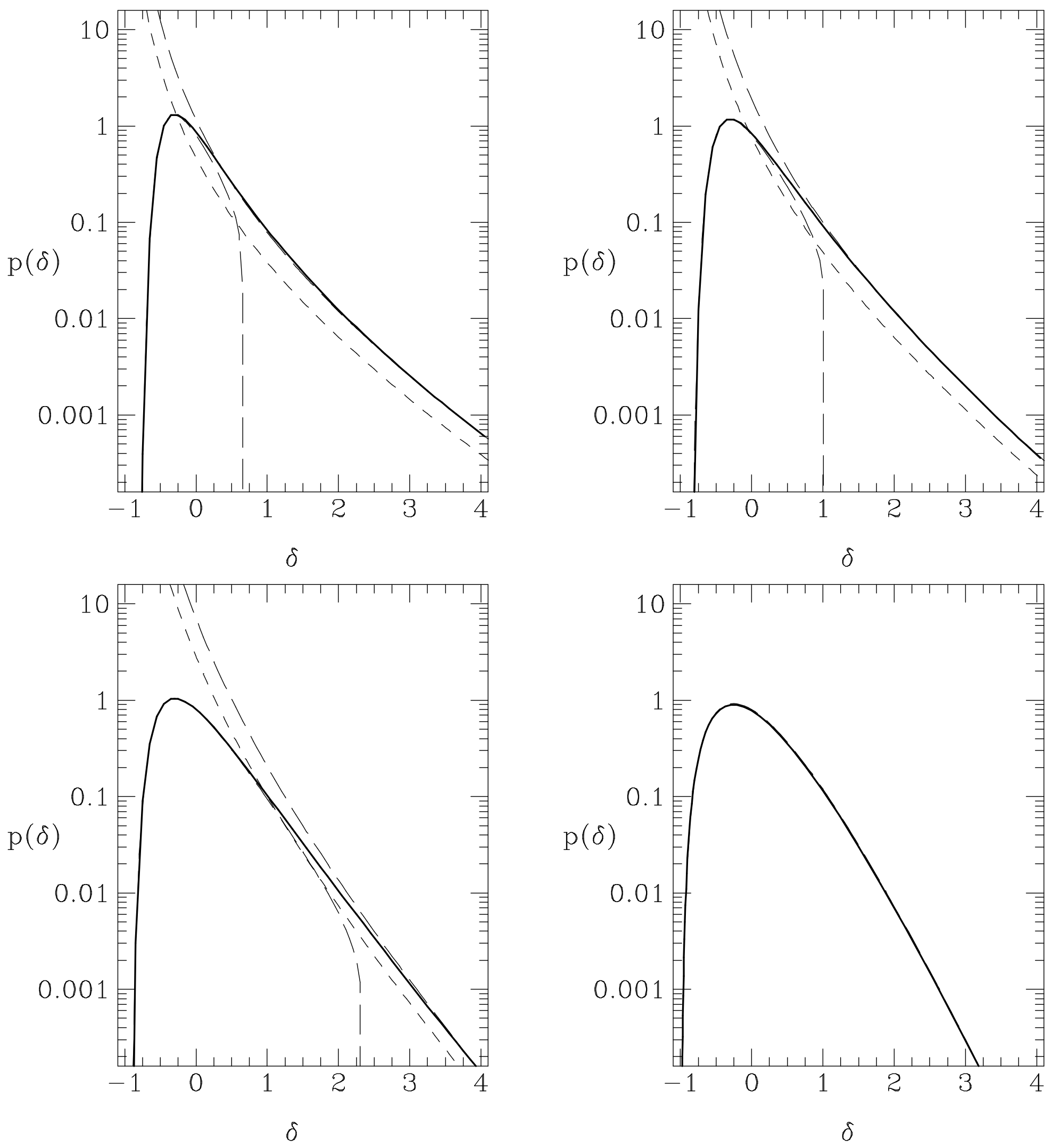

**Fig 3.** The one–point probability distribution function of the density contrast. The thick solid line has been obtained from the equation (18), the long dashed lines correspond to the approximations (47) and (50) and the short dashed line to the approximation (48). From top to bottom and left to right the values of $n$ are successively -3, -2, -1 and 0 and $\sigma = 0.5$ in all cases.

There are mainly two different approximations for the density distribution function depending on the value of $\delta$. When $\delta$ is lower than the critical value $\delta_c$ (that depends on $n$), one can write an expression similar to equation (19),

$$p_1^{\mathcal{S}}(\delta)\mathrm{d}\delta = \frac{1}{-\mathcal{G}_\delta^{\mathcal{S}\,\prime}(\tau)} \left[\frac{1-\tau\mathcal{G}_\delta^{\mathcal{S}\,\prime\prime}(\tau)/\mathcal{G}_\delta^{\mathcal{S}\,\prime}(\tau)}{2\pi\sigma^2}\right]^{1/2} \exp\left(-\frac{\tau^2}{2\sigma^2}\right)\mathrm{d}\delta, \quad \mathcal{G}_\delta^{\mathcal{S}}(\tau) = \delta\,. \tag{47}$$

When the density is greater the $\delta_c$ another approximation can be used that is valid in the large density limit. The behavior of the density distribution in such a case in driven by the shape of the function $\varphi^{\mathcal{S}}(y)$ around its singularity and it reads (Balian & Schaeffer 1989, Bernardeau 1992, Eq.

[49]),

$$p_2^{\mathcal{S}}(\delta)\mathrm{d}\delta = \frac{3a_s\sigma}{4\sqrt{\pi}}(1+\delta)^{-5/2}\exp\left[-|y_s|\delta/\sigma^2+|\varphi_s|/\sigma^2\right]\mathrm{d}\delta. \qquad (48)$$

When $n>0$ the expression (47) is valid whatever the value of $\delta$, otherwise the expression (48) shows that the large density cut-off is in $\exp(-|y_s|\delta/\sigma^2)$ which is very different from a Gaussian like cut-off.

In Fig. 3, I present these approximations compared to an exact integration of the equation (18). Note that the algorithm for the integration in the complex plane roughly follows the same kind of distinction for large and small values of $\delta$. The basic idea for such an algorithm is to define a path for $y$ in the complex plane so that $\varphi(y)-\delta y$ is always a positive real number thus assuring a rapid cutoff of the integral in (18). The value of $y$ to be chosen for the intersection of the path and the real axis is either given by the saddle point $y=-\tau/\mathcal{G}_\delta^{\mathcal{S}\,'}(\tau)$, $\mathcal{G}_\delta^{\mathcal{S}}(\tau)=\delta$ when $\delta<\delta_c$ or by the singular point $y_s$ when $\delta>\delta_c$.

General features can then be presented at this stage. From the form (47) one can infer the form of the cut–off at low density. We obtain,

$$p^{\mathcal{S}}(\delta)\mathrm{d}\delta = \frac{1-n}{4(2\pi\sigma^2)^{1/2}}\left[\frac{2(4-n)}{1-n}\right]^{1/2}(1+\delta)^{\frac{(n-7)}{6}}\exp\left[-\frac{9}{8}\frac{(1+\delta)^{\frac{-(1-n)}{3}}}{\sigma^2}\right]\mathrm{d}\delta. \qquad (49)$$

It shows that the cut-off is sharper when $n$ is small. On the other hand, table 1 shows that the cut-off is more gentle in the high density limit when $n$ is small. In other words, the distribution is more Gaussian like when $n$ is large. This is not a surprising result knowing the dependence of the $S_p$ parameters that are all decreasing functions of $n$ (see Fig. 1).

As can be seen in Fig. 3, the form (47) is extremely accurate for $\delta\lesssim\delta_c$ whereas the form (48) is not very accurate for large densities. It is then of interest to estimate the first corrective term of this form when the large density limit is released. Actually such a term would contain an extra factor $\delta^{-1/2}$ compared to the dominant term. I then propose an approximation based on the form (48) that is more accurate for large values of $\delta$. In such a case, the density distribution can be written,

$$p_3^{\mathcal{S}}(\delta) = \left[1+2\,(0.8-\sigma)\sigma^{-1.3}\rho^{-0.5}\right]p_2^{\mathcal{S}}(\delta). \qquad (50)$$

The corrective factor has been constructed empirically from the shape of the probability distribution function calculated numerically from the form (18). It is quite accurate as long as $\sigma\lesssim 1$. It is possible to get a reasonably good analytical fit of the shape of the distribution function by an interpolation between the form (47) and the form (50). for instance we can use

$$\begin{aligned} p^{\mathcal{S}}(\delta) \approx{}& p_1^{\mathcal{S}}(\delta)+I(\delta)\left[p_3^{\mathcal{S}}(\delta)\exp\left[0.05(1+\delta)^{-4}\right]-p_1^{\mathcal{S}}(\delta)\right] \\ I(\delta) ={}& \frac{\exp(p)}{\exp(p)+\exp(-p)},\quad p=(\delta-\delta_0)\frac{\sigma_0}{\sigma}. \end{aligned} \qquad (51)$$

In the form (51), the factor $\exp[0.05(1+\delta)^{-4}]$ has been put just to regularize this form for small values of $\delta$. The values of $\delta_0$ and $\sigma_0$ can be adjusted to get a smooth interpolation between the two forms. For instance we can take $\delta_0=0.2, 0.4, 2$, $\sigma_0=2, 1.5, 0.8$ for respectively $n=-3,-2,-1$.

Fig. 2 also shows one striking feature for the $n=-1$ case. It turns out that for values of interest for the density and when $\sigma$ is lower than unity, the probability distribution function is quite close to a lognormal distribution

$$p_{\mathrm{l}}(\delta)\mathrm{d}\delta = \frac{1}{(2\pi\sigma_{\mathrm{l}}^2)^{1/2}}\exp\left[-\frac{\left(\ln[1+\delta]+\sigma_{\mathrm{l}}^2/2\right)^2}{2\sigma_{\mathrm{l}}^2}\right]\frac{\mathrm{d}\delta}{1+\delta},\quad \sigma_{\mathrm{l}}^2=\ln(1+\sigma^2). \qquad (52)$$

This distribution was proposed by Coles & Jones (1991) as a good empirical fit for the density probability distribution function. This form, however, does not exhibit the same cut-offs both at large densities and small densities as the distribution functions obtained with the form (18) (Eqs. [48, 49]). That the lognormal distribution is close to the exact behavior when $n = -1$ seems to be a mere coincidence (see Bernardeau & Kofman 1994 for more detailed discussions on this subject).

*3.5. Comparisons with numerical simulations*

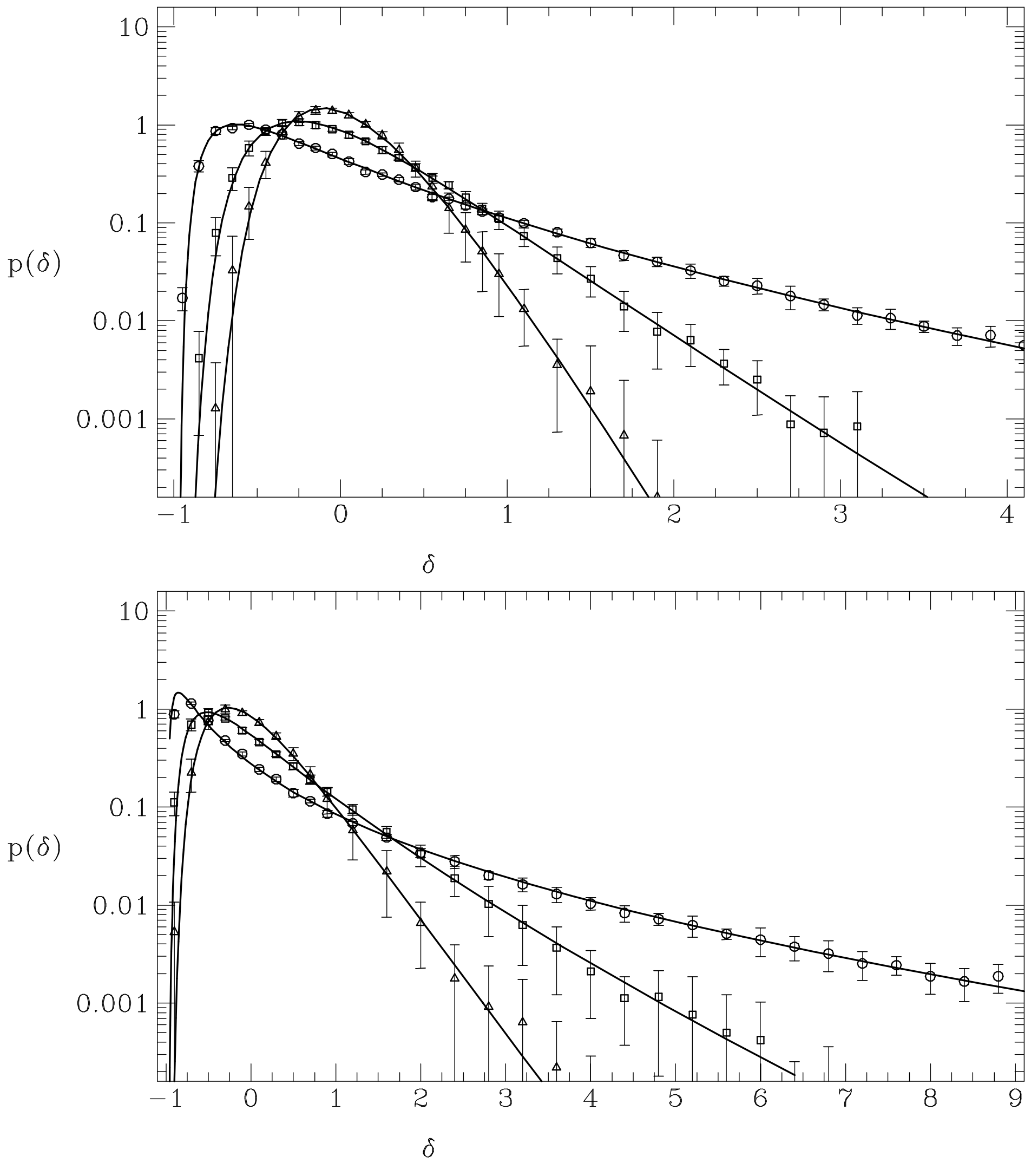

**Fig 4.** Comparison with numerical simulations for the one–point probability distribution function of the density contrast. The upper panel corresponds to the timestep $a/a_0 = 0.6$ and the lower panel to the timestep $a/a_0 = 1$ of a numerical simulation with CDM initial conditions. The density probability distribution function has been measured at three different radius, $R = 5$ h$^{-1}$Mpc (circles), $R = 10$ h$^{-1}$Mpc (squares), $R = 15$ h$^{-1}$Mpc (triangles) in each case. The properties of the local density fluctuations, $n$ and $\sigma$, are given in table 2 for each theoretical curve.

Juszkiewicz et al. (1993b) performed a series of numerical simulations to test predictions of perturbation calculations both on the density field and on the velocity divergence. The filtering, however, was made by a Gaussian window function instead of a top–hat window function which is likely to change the results. In Fig. 1, I present the theoretical value of $S_3$ obtained by Juszkiewicz et al. (1993a) in case of a Gaussian window function. When $n = -3$ the filtering does not introduce any correction either for a top–hat or for a Gaussian window function and the smoothing effects are still quite comparable when $n$ is lower than -2. But for values of $n$ or the order of $-1$ or greater, the smoothing corrections are less important for a Gaussian window function than for a top–hat. To compare the analytical results with numerical simulations one can then find a sort of effective index which would give the correct skewness with a top–hat filter. For instance, for $n = -1$ the effective index would be $n_{\rm eff.} \approx -1.5$, for $n = 0$, $n_{\rm eff.} \approx -1.2$. This recipe allows to estimate the value of $S_4$ for a Gaussian window function. For $n = -1$ and $n = 0$ we obtain respectively $S_4 \approx 20.1$ and $S_4 \approx 16.2$ in good agreement with the result obtained by Catelan & Moscardini (1993) with Monte Carlo numerical integrations. The theoretical probability distribution functions calculated with the effective indices prove to be extremely accurate to predict the shape of the measured probability density distribution functions presented in the paper of Juszkiewicz et al. (1993b).

I also took advantage of a simulation with CDM initial conditions, kindly provided by H. Couchman (Couchman 1991), to compare the theoretical predictions with a realistic model. The simulation contains 2.1 $10^6$ particles, has CDM initial conditions, $H_0 = 50$ km s$^{-1}$ Mpc$^{-1}$, $\Omega = 1$ and was performed in a cubic box of 200 $h^{-1}$ Mpc side with an adaptive P$^3$M code. The measurements of the shape of the probability distribution function of the density contrast have been made at two different timesteps corresponding to an expansion factor of $a/a_0 \approx 0.6$ and $a/a_0 = 1$ where $a_0$ is the final expansion factor (at the end of the simulation we have $\sigma_8 = 0.97$).

The values of the local indices $n$ can be determined from the shape of the initial spectrum for each given smoothing radius. I then approximate the scale dependence of the rms density fluctuation at each smoothing radius by a power law given by the local index. For each timestep I measured the counts in cells probabilities on a grid of $50^3$ points for three different cell radius. For such radius greater then 5 $h^{-1}$Mpc the shot noise is negligeable in this simulation so that the counts in cells give the shape of the numerical density probability distribution function. The rms density fluctuation $\sigma$ is measured in each case. The values obtained for $n$ and $\sigma$ are given in table 2. Knowing $n$ and $\sigma$, it is then possible to calculate the expected shape of the probability distribution function of the density contrast and to compare it to the numerical results. These comparisons are presented in Fig. 4. They show a perfect agreement between the theoretical predictions and the measurements in the simulations. The error bars have been obtained by dividing the simulations in 8 parts and by calculating the rms fluctuation between the different subsets for each smoothing radius and density contrast. These errors are dominated by finite size effects and are thus certainly overestimated for the measurements in the total sample. Perturbation theory proves to be extremely accurate even when $\delta$ is by far larger than unity, and for values of $\sigma$ slightly above unity.

## 4. The smoothing effects for the velocity field

In this part I consider the properties of the probability distribution function of the large–scale velocity divergence field, $\theta(\mathbf{x})$ (Eq. [1]).

### *4.1. The generating function of the cumulants with the smoothing effects*

Let me comment a bit more on the analytical derivations presented in the previous part for the density field. The central argument is that the function $\mathcal{G}_\delta$ obtained without taking into account the smoothing effects is exact for a given mass scale. Then the resulting expression of the function $\mathcal{G}_\delta^{\mathcal{S}}(\tau)$ is simply given by the function $\mathcal{G}_\delta(\tau)$ but calculated at the mass, $M_0$, of interest. The relevant value of $\tau$ depends on the mass scale through the function $\sigma$ in a way that can be deduced from the equation (15) by identifying the terms under the exponential in both Lagrangian and Eulerian case (which gives $\tau^{\mathcal{S}}/\sigma(M_0)=\tau/\sigma(R_0)$). The relationship between the mass $M_0$ and the scale $R_0$ is directly related to the value of $\mathcal{G}_\delta^{\mathcal{S}}(\tau^{\mathcal{S}})$ (see again Eq. [15] that gives $1+\mathcal{G}_\delta^{\mathcal{S}}(\tau)=3M_0/(4\pi R_0^3)$). Then we obtain the equation (38).

A similar analysis can be done for the cumulants of the divergence of the velocity field. They are not affected by smoothing in Lagrangian space for a given mass scale (this result can be obtained in a similar way than for the density field since the expression of the velocity divergence as a function of the displacement field involves the same geometrical expressions). The generating function of the vertices for the divergence can then simply be calculated at the mass scale of interest, that is at the same point $\tau$ as previously,

$$\mathcal{G}_\theta^{\mathcal{S}}(f(\Omega,\Lambda)\tau^{\mathcal{S}})=\mathcal{G}_\theta\left[f(\Omega,\Lambda)\tau^{\mathcal{S}}\ \frac{\sigma\left([1+\mathcal{G}_\delta^{\mathcal{S}}(\tau^{\mathcal{S}})]^{1/3}R_0\right)}{\sigma(R_0)}\right]. \tag{53}$$

With (18) and (45) it gives the expression of $\mathcal{G}_\theta^{\mathcal{S}}$ as a function of $\tau$.

*4.2. Examples of cumulants*

The expression of the generating function of the coefficients $T_p$ can naturally be derived from the expression of $\mathcal{G}_\theta^{\mathcal{S}}$,

$$\varphi_\theta^{\mathcal{S}}(a,y)=y\mathcal{G}_\theta^{\mathcal{S}}(a,\tau)-\frac{1}{2}y\tau\frac{\mathrm{d}}{\mathrm{d}\tau}\mathcal{G}_\theta^{\mathcal{S}}(a,\tau),\quad \tau=-y\frac{\mathrm{d}}{\mathrm{d}\tau}\mathcal{G}_\theta^{\mathcal{S}}(a,\tau). \tag{54}$$

The expansion of this function with respect to $y$ then can give the expressions of the coefficient $T_p$ as a function of the logarithmic derivatives of $\sigma$. I give here their expression for an Einstein-de Sitter Universe,

$$\begin{aligned}
-T_3 &= \frac{26}{7}+\gamma_1,\\
T_4 &= \frac{12088}{441}+\frac{338\,\gamma_1}{21}+\frac{7\,{\gamma_1}^2}{3}+\frac{2\,\gamma_2}{3},\\
-T_5 &= \frac{94262120}{305613}+\frac{161440\,\gamma_1}{567}+\frac{260\,{\gamma_1}^2}{3}+\frac{235\,{\gamma_1}^3}{27}+\frac{130\,\gamma_2}{7}+\frac{50\,\gamma_1\,\gamma_2}{9}+\frac{10\,\gamma_3}{27},\\
T_6 &= 4694+5951\,\gamma_1+2802\,{\gamma_1}^2+581.2\,{\gamma_1}^3+44.81\,{\gamma_1}^4+480.3\,\gamma_2\\
&\quad +297.1\,\gamma_1\,\gamma_2+45.56\,{\gamma_1}^2\,\gamma_2+3.889\,{\gamma_2}^2+15.82\,\gamma_3+4.815\,\gamma_1\,\gamma_3+0.1852\,\gamma_4,\\
-T_7 &= 90310+146000\,\gamma_1+93680\,{\gamma_1}^2+29800\,{\gamma_1}^3+4704\,{\gamma_1}^4+294.8\,{\gamma_1}^5\\
&\quad +13380\,\gamma_2+12710\,\gamma_1\,\gamma_2+3991\,{\gamma_1}^2\,\gamma_2+414.8\,{\gamma_1}^3\,\gamma_2+284.1\,{\gamma_2}^2+88.15\,\gamma_1\,{\gamma_2}^2\\
&\quad +562.9\,\gamma_3+351.5\,\gamma_1\,\gamma_3+54.44\,{\gamma_1}^2\,\gamma_3+7.778\,\gamma_2\,\gamma_3+11.23\,\gamma_4+3.457\,\gamma_1\,\gamma_4+0.08642\,\gamma_5,\\
&\dots
\end{aligned}$$

These functions are presented in Fig. 1 for power law spectra. Note that the first two have already been derived with direct perturbative calculations (Bernardeau 1994b). In general the dependence of the coefficients $T_p$ with the cosmological parameters is simple and can be easily obtained from the dependence of $\mathcal{G}_\theta$ with the cosmological parameters. We then approximately have

$$T_p(\Omega,\Lambda)\approx\frac{1}{f(\Omega,\Lambda)^{p-2}}T_p(\Omega=1,\Lambda=0). \tag{55}$$

The use of the $\Omega$ dependence exhibited by these coefficients has already been proposed to determine $\Omega$ from the large–scale velocity fields (Bernardeau et al. 1994, Bernardeau 1994b). This dependence will also give a strong $\Omega$ dependence for the shape of the probability distribution function.

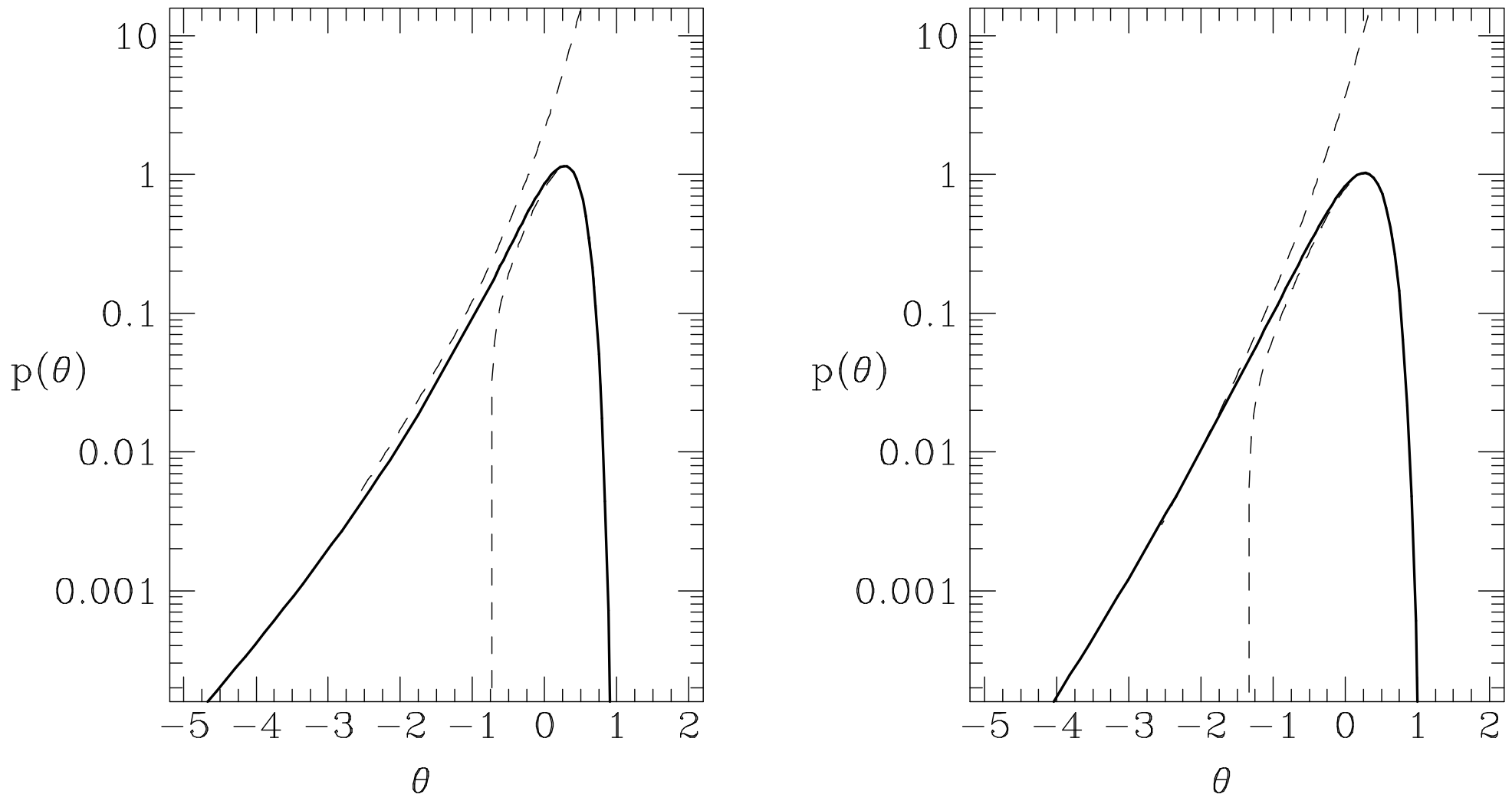

**Fig 5.** The one–point probability distribution function of the divergence of the velocity field for $n = -3$ (left panel) and $n = -2$ (right panel). The thick solid line is the exact result of the quasilinear regime for an Einstein–de Sitter Universe. The short dashed lines correspond to the approximations (51) and (58).

*4.3. The shape of the probability distribution function*

A general rule can be given for the dependence of the probability distribution function with the cosmological parameters. It takes advantage of the dependence of the cumulants with $\Omega$ and $\Lambda$ (Eq. [50]) that leads to,

$$p_\theta^{\mathcal{S}}(f(\Omega,\Lambda),\theta,\sigma_\theta)\mathrm{d}\theta = p_\theta^{\mathcal{S}}\left[1,\frac{\theta}{f(\Omega,\Lambda)},\frac{\sigma_\theta}{f(\Omega,\Lambda)}\right]\frac{\mathrm{d}\theta}{f(\Omega,\Lambda)}. \tag{56}$$

**Table 2.** The parameters of the critical point as a function of the spectral index, $n$, for the distribution of the velocity divergence.

| $n$ | $\theta_c$ | $y_{s\theta}$ | $a_{s\theta}$ | $\varphi_{s\theta}$ |
|---|---|---|---|---|
| $-3$ | $-0.750$ | $-0.222$ | 1.67 | $-0.042$ |
| $-2.5$ | $-0.961$ | $-0.266$ | 1.99 | $-0.061$ |
| $-2$ | $-1.348$ | $-0.332$ | 2.51 | $-0.100$ |
| $-1.5$ | $-2.325$ | $-0.447$ | 3.63 | $-0.201$ |

The probability distribution function can then be calculated exactly the same way as the one of the density contrast, Fig 5. Once again two different analytical approximations can be derived. One is valid when the divergence is greater than the critical value $\theta_c$ (depending on $n$, see table 2) and it is given by

$$p_{1\theta}^{\mathcal{S}}(\theta)\mathrm{d}\theta = \frac{1}{-\mathcal{G}_\theta^{\mathcal{S}}{}'(\tau)}\left[\frac{1-\tau\mathcal{G}_\theta^{\mathcal{S}}{}''(\tau)/\mathcal{G}_\theta^{\mathcal{S}}{}'(\tau)}{2\pi\sigma_\theta^2}\right]^{1/2}\exp\left(-\frac{\tau^2}{2\sigma_\theta^2}\right)\mathrm{d}\theta,\quad \mathcal{G}_\theta^{\mathcal{S}}(\tau)=\theta\,. \tag{57}$$

The other one is valid when the divergence is lower than the critical value,

$$p^{\mathcal{S}}_{2\theta}(\theta)\mathrm{d}\theta = \frac{3a_{s\theta}\sigma_\theta}{4\sqrt{\pi}}(3/2-\theta)^{-5/2}\exp\left[|y_{s\theta}|\theta/\sigma_\theta^2+|\varphi_{s\theta}|/\sigma_\theta^2\right]\mathrm{d}\theta, \qquad (58)$$

where the parameters $a_{s\theta}$, $y_{s\theta}$ and $\varphi_{s\theta}$ are given in table 2 for few values of $n$.

It appears that when $n \geq -1$ there is no singular point for the function $\varphi^{\mathcal{S}}_\theta(y)$ so that the expression (57) is always valid.

The form (57) is extremely accurate for large values of $\theta$ whereas the form (58) underestimates the exact derivation of the probability distribution function. I then propose a correction to the form (58) which is in a better agreement with the exact form,

$$p^{\mathcal{S}}_{3\theta}(\theta)\mathrm{d}\theta = \left[1+30(.8-\sigma_\theta)\sigma_\theta^{-1.3}(1.5-\theta)^{-1.5}\right]p^{\mathcal{S}}_{2\theta}(\theta)\mathrm{d}\theta. \qquad (59)$$

In Fig. 5, the shape of these approximative forms are shown. Note that these forms have been given for an Einstein–de Sitter Universe but can be easily extended to any cosmological model with equation (56).

The case $n = -1$ is worth investigating for two reasons. First it corresponds roughly to the observed value of the power law index at large scales. Moreover, it is possible to derive an analytical form that fits perfectly well the exact numerical derivation. This approximation is based on the approximate form (14) for the function $\mathcal{G}_\delta$. One can then show that $\mathcal{G}^{\mathcal{S}}_\theta(\tau) = \tau(1+\tau^2/9/f^2(\Omega,\Lambda))^{1/2} - \tau^2/3/f(\Omega,\Lambda)$. The calculation of the expression (57) then leads to the expression,

$$p^{\mathcal{S}}_\theta(\theta)\mathrm{d}\theta = \frac{([2\kappa-1]/\kappa^{1/2}+[\lambda-1]/\lambda^{1/2})^{-3/2}}{\kappa^{3/4}(2\pi)^{1/2}\sigma_\theta}\,\exp\left[-\frac{\theta^2}{2\lambda\sigma_\theta^2}\right]\mathrm{d}\theta, \qquad (60)$$

with

$$\kappa = 1+\frac{\theta^2}{9\lambda f^2(\Omega,\Lambda)}, \qquad \lambda = 1-\frac{2\theta}{3f(\Omega,\Lambda)}. \qquad (61)$$

*4.4. Comparison with numerical simulations*

To obtain a smoothed density field from a discrete sample of points is a simple task. To obtain the smoothed velocity divergence is not so straightforward and it is worth noting that the smoothing is supposed to be done with a volume-weighted scheme, that is the velocities, at any point of the considered volume, should be equally weighted, regardless of the local density.

I used the CDM numerical simulation once again to check the shape of the expected distribution function of $\theta$. A grid of $40^3$ cubic cells (of size 5 $h^{-1}$Mpc) have been used to define the velocity field in the whole simulation, the velocity associated with each cell being the average of the velocities of the points contained in the cell. A spherical box of a given radius has been centered on each cell. The smoothed velocity in each box is obtained by taking the average of the velocities in the cells that are part of the box (weighted by the fraction of the volume of the cell actually contained in the box). This procedure insures that the filtering has been made with a volume weighted scheme, as soon as the ratio of the radius of the box by the size of the cells is large enough. I made the measurement only for a soothing radius of 10 $h^{-1}$Mpc and 15 $h^{-1}$Mpc to keep a high enough ratio (and only for $a/a_0 = 1$). The divergence has been calculated with finite differences using the initial grid.

The resulting probability distribution functions of the velocity divergence are presented in fig. 6. They appear to be extremely noisy in the tails but the central bump is perfectly well described by the analytical form (60) with $\Omega = 1$. The values of $\sigma_\theta$ to be used in (60) can be determined by suppressing the noisy part of the tails. They are respectively $\sigma_\theta = 0.45$ and $\sigma_\theta = 0.32$. The differences between $\sigma$ and $\sigma_\theta$ (see table 2) for the same smoothing radius are thought to be due to the smoothing scheme used for the velocity field. Indeed, it tends to smooth the velocity field in an effective radius greater than the size of the box (especially because of the finite differences), and thus to reduce the value of $\sigma_\theta$. Moreover the noise is particularly important in the underdense regions (for which the velocity is poorly known) and give birth to noisy data in both the low and the high density tails (as can be seen with an analysis of the $\theta$–$\delta$ correlations). For instance for $R_0 = 10$ $h^{-1}$Mpc about 2% of the measured divergences are in the tails (defined by $\theta < -2$ or $\theta > 1$) while 0.6% of the cells of the grid were completely empty (in such a case the velocity of the cell was taken as the average of the velocities in the surrounding cells). Note also that when the tails are suppressed the values found for $T_3$ and $T_4$ are respectively $-1.8 \pm 0.15$ and $4.9 \pm 1.5$ in good agreement with the theoretical predictions for $n = -1$ and $\Omega = 1$ ($-1.7$ and $4.6$).

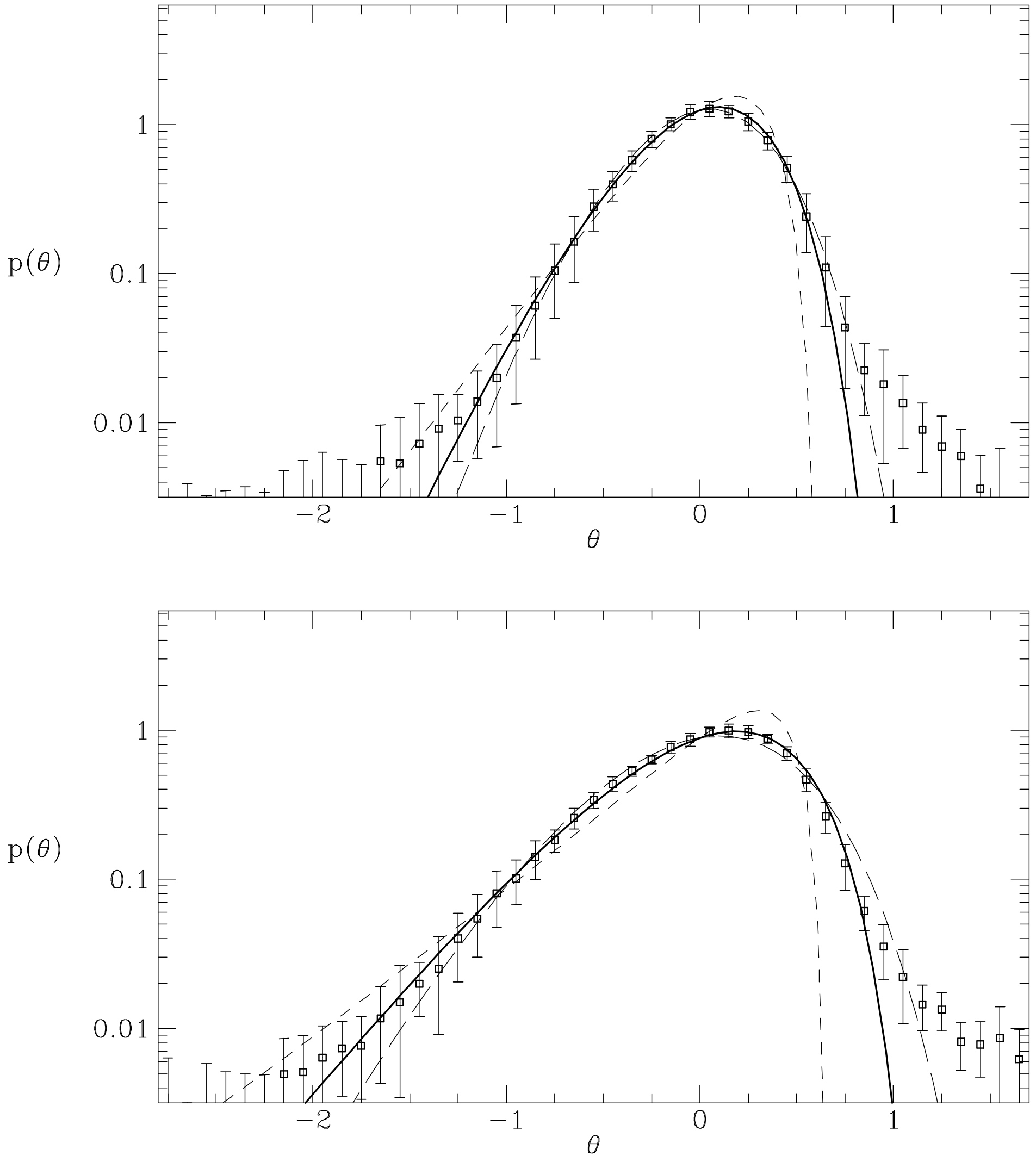

**Fig 6.** The one–point probability distribution function of the velocity divergence obtained in an $\Omega = 1$ CDM numerical simulation. The smoothing radius are 15 $h^{-1}$Mpc (top panel) and 10 $h^{-1}$Mpc (bottom panel) and $\sigma_\theta$ is respectively 0.32 and 0.45. The solid lines correspond to the formula (60) for $\Omega = 1$, the dashed lines for $\Omega = 0.3$ and the long dashed lines for $\Omega = 3$.

The form (60) proves also to be extremely accurate to give the shape of the distribution function of the velocity divergence obtained by Juszkiewicz et al. (1993b).

When the data are compared to the expected shape of the probability distribution function of $\theta$ for other values of $\Omega$, a clear discrepancy can be seen, especially for low values of $\Omega$ (short dashed lines in Fig. 6). Such a function could then be of crucial interest to constrain observationally the value of $\Omega$.

## 6. Conclusion

In this paper, I have proposed a method to calculate the cumulants of the large–scale distribution function of the density and the velocity divergence when a top hat smoothing window function is applied to the fields. The generating functions of the cumulants (defined in Eq. [10]) are given by the equations (41) and (54) as a function of the shape of the power spectrum. The probability distribution functions can then be obtained from the form (18) used with the appropriate generating function $\varphi(y)$. This form has been derived assuming that the ratios $\left\langle\delta^p\right\rangle_c/\left\langle\delta^2\right\rangle^{p-1}$ (or $\left\langle\theta^p\right\rangle_c/\left\langle\theta^2\right\rangle^{p-1}$) are well approximated by their large–scale limit. Various approximations for the probability distribution functions have then been given (Eqs. [47, 50, 52] and [57-61]). The $n = -1$ case received a particular focus: it has been shown that the lognormal distribution was particularly close to the exact derivation of the density one–point distribution function (although this distribution has none of the general properties of the exact results, see text) and a simple analytical fit has also been given in that case for the velocity divergence distribution function (Eqs. [60-61]).

For fields smoothed by a Gaussian window function I proposed an empirical model that consists in replacing the index of the power spectrum by an effective one that reproduces correctly the skewness. The distribution functions can then be calculated by the results obtained for a top hat window function but using the effective index. This is not, however, an exact result.

Comparisons with numerical simulations proved that these results are extremely accurate in the whole quasilinear regime. It implies that the hypothesis of constant ratios, $\left\langle\delta^p\right\rangle_c/\left\langle\delta^2\right\rangle^{p-1}$, is extremely powerful in such a regime. Results of numerical simulations (Bouchet & Hernquist 1992, Lucchin et al. 1993) indicate that such an hypothesis is no more exact in the fully nonlinear regime when $\sigma$ exceeds unity. It may be, however, interesting to know what features are induced by the results of the large scale cumulant behaviors when they are applied in the nonlinear regime. This is left for a coming paper.

Applications to observational data are also left for another paper (Kofman & Bernardeau, in preparation).

Note: The Fortran program used to integrate Equation (18) is available upon request (email address: fbernard@amoco.saclay.cea.fr).

*Acknowledgments:* The author thanks D. Bond, L. Kofman, R. Juszkiewicz, F. Bouchet for useful discussion and H. Couchman for providing the numerical simulation. Part of this work was done while the author was at the Service de Physique Théorique, CE de Saclay, France